\documentclass[%
 reprint,
superscriptaddress,
 amsmath,amssymb,
 aps
]{revtex4-2}

\usepackage{graphicx}
\usepackage{dcolumn}
\usepackage{bm}
\usepackage{comment}

\begin{document}

\preprint{APS/123-QED}


\title{Atomic diffusion due to hyperatomic fluctuation for quasicrystals}

\author{Yuki Nagai}
\email{nagai.yuki@jaea.go.jp}
\affiliation{%
CCSE, Japan Atomic Energy Agency, 178-4-4, Wakashiba, Kashiwa, Chiba 277-0871, Japan
}
\affiliation{
Mathematical Science Team, RIKEN Center for Advanced Intelligence Project (AIP), 1-4-1 Nihonbashi, Chuo-ku, Tokyo 103-0027, Japan
}
%
\author{Yutaka Iwasaki}%
\email{IWASAKI.Yutaka@nims.go.jp}
\affiliation{%
National Institute for Materials Science
(NIMS), 1-2-1 Sengen, Tsukuba, Ibaraki 305-0047, Japan
}%
\affiliation{%
Department of Advanced Materials Science, The University of Tokyo, 5-1-5 Kashiwanoha, Kashiwa, Chiba 277-8561, Japan
}%


\author{Koichi Kitahara}
\affiliation{%
Department of Materials Science and Engineering, National Defense Academy, 1-10-20 Hashirimizu, Yokosuka, 239-8686, Kanagawa, Japan.
}%

\author{Yoshiki Takagiwa}%
\affiliation{%
National Institute for Materials Science
(NIMS), 1-2-1 Sengen, Tsukuba, Ibaraki 305-0047, Japan
}%

\author{Kaoru Kimura}
\affiliation{%
National Institute for Materials Science
(NIMS), 1-2-1 Sengen, Tsukuba, Ibaraki 305-0047, Japan
}%
\affiliation{%
Department of Advanced Materials Science, The University of Tokyo, 5-1-5 Kashiwanoha, Kashiwa, Chiba 277-8561, Japan
}%
\author{
Motoyuki Shiga
}
\affiliation{%
CCSE, Japan Atomic Energy Agency, 178-4-4, Wakashiba, Kashiwa, Chiba 277-0871, Japan
}%


\date{\today}

\begin{abstract}
A quasicrystal is an ordered but non-periodic structure understood as a projection from a higher dimensional periodic structure. 
Some physical properties of quasicrystals are different from those of conventional solids. 
An anomalous increase in heat capacity at high temperatures has been discussed for over two decades as a manifestation of a hidden high dimensionality of quasicrystals.
A plausible candidate for this origin has been phason, which has excitation modes originating from additional degrees of freedom in the higher-dimensional lattice.
However, most theoretical studies on phasons have used toy models. 
A theoretical study of the heat capacity of realistic quasicrystals or their approximants has yet to be conducted because of the huge computational complexity.
To bridge this gap between experiment and theory, we show experiments and molecular simulations on the same material, an Al--Pd--Ru quasicrystal, and its approximants. 
We show that at high temperatures, aluminum atoms diffuse with discontinuous-like jumps, and the diffusion paths of the aluminum can be understood in terms of jumps corresponding to hyperatomic fluctuations in six-dimensional space. 
It is concluded that the anomaly in the heat capacity of quasicrystals arises from extra degrees of freedom due to hyperatomic fluctuations that play a role in diffusive Nambu--Goldstone modes. 
\end{abstract}

\maketitle

\section*{Introduction}
Quasicrystals (QCs)\cite{Shechtman1984,Levine1984} are solids with quasiperiodic atomic structures. 
 Mathematically, a quasiperiodic structure is constructed by a projection from a higher-dimensional space \cite{Yamamoto1996}: 
 Every quasiperiodic structure in three-dimensional physical space can be described as a projection from a hypothetical higher-dimensional periodic crystal structure called ``hyperlattice''.
The hyperlattice concept is commonly used to explain the static structures of QCs\cite{Boudard1992,Takakura2007}.
 
QCs have unique physical properties that are not observed in conventional solids\cite{Deguchi2012, Kamiya2018, Tamura2021}. 
In particular, several quasicrystals at high temperatures show significant increases in heat capacity. 
The heat capacity per atom at a constant volume $c_V$ for icosahedral Al--Pd--Mn \cite{EDAGAWA2000646,EDAGAWA2001293,Fukushima_2021}, Al--Cu--Fe \cite{Prekul:2008aa}, Al--Cu--Ru \cite{Tamura2021MT-MB2020004}, and decagonal Al--Cu--Co \cite{EDAGAWA2000646,EDAGAWA2001293} QCs show large upward deviations from the Dulong--Petit limit $3k_{\rm B}$, where $k_{\rm B}$ is the Boltzmann constant. 
Such anomalous heat capacity at high temperatures has been debated for over two decades, and a potential connection to the hidden higher-dimensional properties of QCs has been discussed.

The phason, which contains additional degrees of freedom because of the hyperlattice structure \cite{PhysRevLett.54.1520,PhysRevB.34.3345}, has been suggested as the origin of the high-temperature anomalous heat capacity \cite{10.21468/SciPostPhys.9.5.062,Fukushima_2021}.
The phason modes originate from the fact that the free energy of the system is invariant under a rigid translation of the parallel space in the perpendicular dimension, where the hyperlattice decomposes into the parallel (physical) and the perpendicular space. 
In many solids, heat capacity at high temperatures approaches the Dulong--Petit limit $3k_{\rm B}$, because full vibrational-mode degrees of freedom amount to three degrees of freedom per atom, each corresponding to a quadratic kinetic energy term and a harmonic potential energy term. 
However, the vibrations in QCs are beyond harmonic oscillations along the phason degrees of freedom, which may contribute to the increased heat capacity.
Inelastic neutron scattering \cite{Coddens2000} and coherent X-ray scattering \cite{Francoual2003} experiments on Al--Pd--Mn QC have observed characteristic phason excitation and diffusive modes above approximately $700\,\mathrm{K}$. 
These results suggest that phason is excited at high temperatures with anomalous atomic motion, consistent with the temperature range where anomalous heat capacity is observed.
In addition, experimental evidence of anomalous atomic vibrations at specific atomic sites in the structure at $1100\,\mathrm{K}$ has been observed in decagonal Al--Ni--Co QCs \cite{Nature_abe}.

Further evidence that phason may be the cause of the anomalous heat capacity is seen in the relationship between the degree of approximation and the heat capacity in quasicrystalline approximant crystals (ACs), which are periodic crystals that have the same local structure as the corresponding QC. The structure of ACs is classified by the degree of approximation represented by two consecutive numbers in the Fibonacci sequence, such as 1/0, 1/1, 2/1, $\cdots$, $q_n$/$q_{n-1}$.
A larger $q_n$ in an AC corresponds to a larger unit cell and a structure that is closer to that of a QC.
As the degree of approximation increases, the AC becomes closer in structure to the QC.
The AC-specific heat is expected to become more anomalous as it approaches the QC limit because the phason degrees of freedom may increase\cite{Tamura2021MT-MB2020004,Fukushima_2021}.

So far, most theoretical studies on phasons have used toy models consisting of a single atomic species in one or two dimensions \cite{Engel2007,Engel2010}. 
The three-dimensional atomic motion of multiple-element QCs has been studied experimentally in decagonal phases such as Al--Ni--Co QCs \cite{Nature_abe}. 
In a recent molecular dynamics (MD) simulation by Mihalkovi\v{c} and Widom \cite{Mihalkovic2020}, an AC of Al--Cu--Fe quasicrystals containing 9846 atoms was studied, focusing on its energetic stability. 
However, it is not clear what contributes to the high-temperature anomalous heat capacity in actual materials, although phason is a plausible candidate.
Additionally, a direct computation of heat capacity has never been done before because to adequately describe the dynamic behavior of QCs accounting for a huge number of atoms and complex atomic interactions is required.

To bridge this large gap between experiment and theory, 
we study the same material from both theoretical and experimental approaches.
First, we synthesized Al--Pd--Ru icosahedral QC and its ACs and observed the high-temperature anomalous heat capacity. 
We then performed a machine-learning molecular dynamics (MLMD) simulation for Al--Pd--Ru ACs and qualitatively reproduced the increase of the heat capacity.
We herein show that Al atoms diffuse with discontinuous jumps at the temperature range where anomalous heat capacity is observed.
%
The diffusion path of Al atoms can be understood in terms of hyperatomic fluctuations in six-dimensional space. 
Considering this atomic diffusion due to hyperatomic fluctuations to be the ``phason" diffusion, this diffusion can be understood as a diffusive Nambu-Goldsone mode.
Based on this observation, we conclude that atomic diffusion due to hyperatomic fluctuations in hyperdimensional space, which consists of a series of discontinuous atomic jumps, is the origin of the high-temperature anomalous heat capacity.

\section*{Heat capacity: experiment and simulation} \label{sec:specificheat}

 \subsection*{Experiment}
\begin{figure*}[th]
  \begin{center}
\includegraphics[keepaspectratio,width=1 \linewidth, page=1]{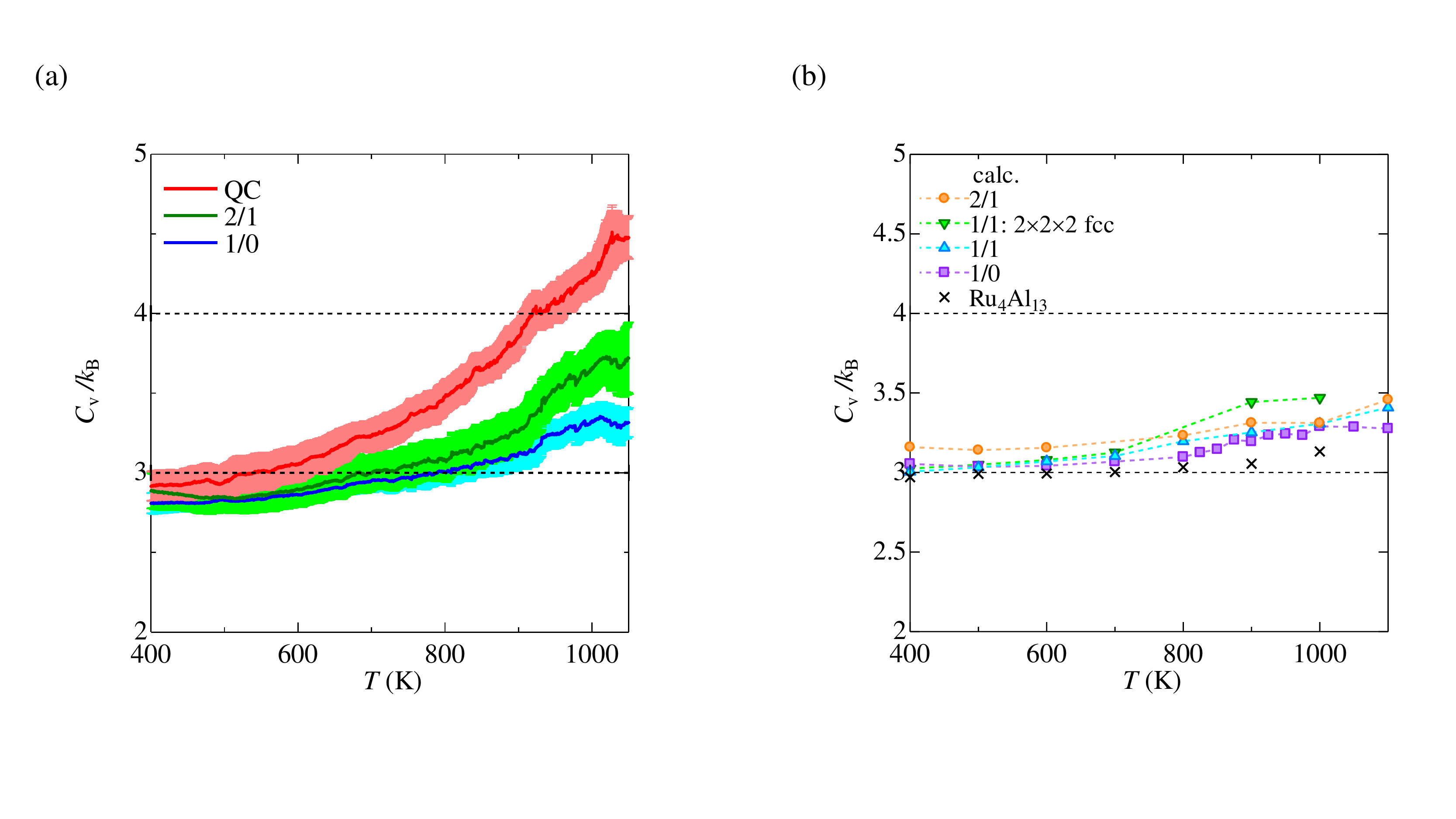}
    \caption{\label{fig:cv_ex} (Color online) Temperature dependence of heat capacity at constant volume normalized by Boltzmann constant $C_{V}/k_{\mathrm{B}}$ (a) in the experiment for Al--Pd--Ru QC (red), 2/1 AC (green), and 1/0 AC (blue), and (b) in the calculation for 2/1 AC (orange circle), 1/1 AC (blue triangle), 1/1 AC with $2\times2\times2$ face-centered cubic primitive supercell (green triangle), 1/0 AC (purple square), and Ru$_4$Al$_{13}$ (cross), where the thick light-colored line on each measured line stands for the standard deviation.}
    \end{center}
\end{figure*}

 Fig.~\ref{fig:cv_ex}(a) shows the experimental results of the constant-volume heat capacity, $C_{V}$, of Al--Pd--Ru icosahedral QC and its ACs.
It can be seen that $C_{V}$ largely deviates from Dulon--Petit’s law, 3$k_{\mathrm{B}}$ with increasing temperature.

The deviation from the $3k_{\mathrm{B}}$ is the largest for QCs, followed by the 2/1 AC and 1/0 AC.
Thus, the $C_{V}$ anomaly of the ACs increases as the AC structure becomes similar to the QC.

This systematical trend is consistent with previous work \cite{Fukushima_2021} that reported $C_{V}$ values for various aluminum--transition metal QCs and ACs.

\subsection*{Simulation: machine-learning molecular dynamics }
  
The machine-learning molecular dynamics (MLMD) simulations of Al--Pd--Ru ACs are conducted under periodic boundary conditions.
We used a potential energy function from an artificial neural network (ANN) that imitates the Born-Oppenheimer energies obtained by first-principles density-functional theory (DFT). 
The ANN was trained by a self-learning hybrid Monte Carlo (SLHMC) method\cite{PhysRevB.102.041124,doi:10.1063/5.0055341} with a combination of \texttt{PIMD}\cite{PIMD, PhysRevB.102.041124},  the Vienna Ab initio Simulation Package (\texttt{VASP}) \cite{PhysRevB.47.558,PhysRevB.54.11169}
and Atomic Energy Network (\texttt{aenet}) \cite{ARTRITH2016135} software.
The computational details are provided in the Supporting Information.

We show the heat capacity directly calculated by the ensemble average of the energy fluctuation as $C_V = \langle (\Delta E)^2 \rangle/(k_{\mathrm{B}}T^2)$. 
Figure \ref{fig:cv_ex}(b) shows the calculated heat capacity for Al$_{13}$Ru$_4$, 1/0, 1/1, and 2/1 Al--Pd--Ru ACs. 
The simulation of Al$_{13}$Ru$_4$ reproduces the conventional Dulong–-Petit limit, 3$k_{\mathrm{B}}$, as observed in experiments \cite{Tamura2021MT-MB2020004}.
For ACs, we found that the calculated heat capacity depends on temperature and becomes larger than the Dulong--Petit limit at high temperature. 
Here, the difference between 1/1 AC and 1/1 AC with $2\times2\times2$ face-centered cubic primitive cells in Fig.~\ref{fig:cv_ex}(b) is the difference of the number of atoms in a unit cell, 128 and 256, respectively.
There is some system size dependence in the simulated results for the 1/1 AC case. 
To make a qualitative comparison with experiment, one needs to add more atoms in the unit cell, although such a computation is too expensive at present.

\section*{Atomic structure and phasons} \label{sec:atomicstructure}
The local atomic structures of the QCs and ACs are similar, the main difference being how the cluster structures are connected. 
Therefore, the cause of the anomalous heat capacity in QCs can be revealed by calculating the corresponding ACs.
To understand high-temperature anomalous heat capacity, we introduce the crystal structure of ACs in the Al--Pd--Ru system.

\begin{figure*}[tbh]
 \begin{center}
\includegraphics[keepaspectratio,width=0.9 \linewidth]{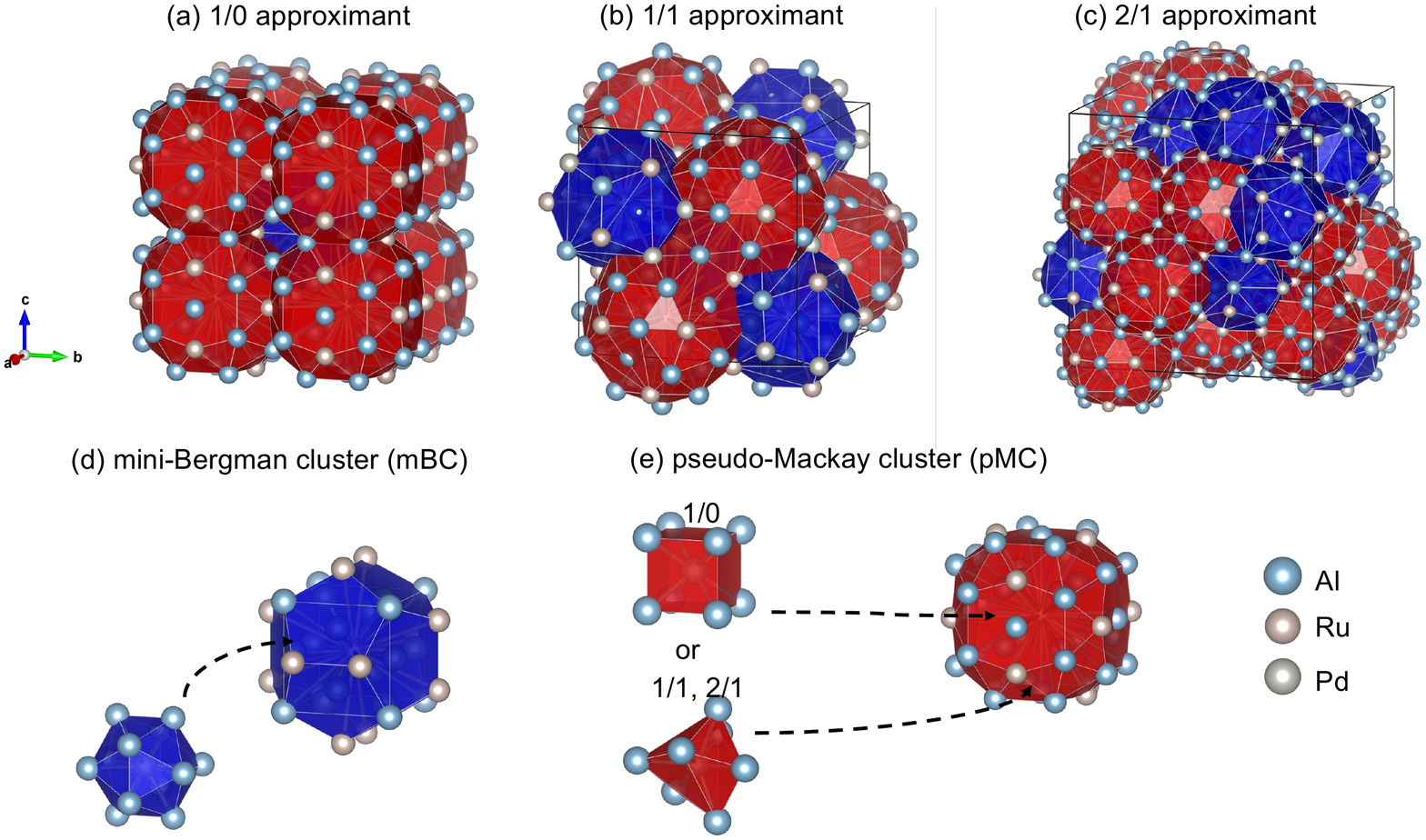}

\caption{\label{fig:structure} (Color online) Atomic structures of Al--Pd--Ru ACs, (a): 1/0 AC, (b): 1/1 AC, (c): 2/1 AC, (d): mini-Bergman cluster and (e): the pseudo-Mackay cluster of each AC. Here, the mini-Bergman and pseudo-Mackay clusters have an inner shell.}
\end{center}
\end{figure*}

Figure~\ref{fig:structure} shows the 1/0 AC, 1/1 AC, and 2/1 AC model structures generated by the modified Katz--Gratias--Boudard model of the Al--Pd--Ru system\cite{Krajci1995,Krajci2003,Krajci2007}. 
The structure of Al--Pd--Ru ACs is described as a dense packing of two types of clusters\cite{Fujita2013}, the so-called mini-Bergman and pseudo-Mackay clusters shown in Fig.~\ref{fig:structure}(d) and (e).
These structures were obtained assuming the smallest size of the hyperatoms in six dimensions, called "the occupation domain". 
Note, however, that these model structures are not necessarily energetically favorable and may not represent the actual structure of the AC at finite temperatures. 
The actual structural ensemble in thermal equilibrium can only be obtained after MD simulations. 
In the next section, we will show the results of the MD simulations starting from these model structures.

\section*{MD analysis} \label{sec:realtime}

\subsection*{Atom jumping}
MD simulations of Al--Pd--Ru AC show several energetically favorable structures apart from the prototype shown in Fig.~\ref{fig:structure}. 
Those structures are different at the Al sites. 
It was found that some Al atoms are mobile in the ACs and randomly jump from one site to another during MD simulations in thermal equilibrium. 
This jump is detected at high temperatures where the heat capacity becomes anomalous.

Figures \ref{fig:phason}(a)--(c) show the trajectories of some Al atoms for 1/1 AC at a temperature of 1000 K. 
Pd, Ru, and some Al atoms composing the inner shell of the mini-Bergman cluster are immobile and oscillate at their respective positions, as shown in Fig.~\ref{fig:phason}(a). 
However, some other Al atoms composing the inner shell and edge of the pseudo-Mackay cluster are moving with almost discontinuous jumps, as shown in Fig.~\ref{fig:phason}(b)--(c). 
This may correspond to what was predicted as "phason flips" in the QC model system \cite{Engel2010,Widom2008}, but detecting atomic jumps in real-time from MD simulations is a new finding. 
At 1000~K, the atomic jumps occur at about 100 $\mathrm{ps}$, which is much longer than conventional phonon oscillations; an analysis of trajectories up to 2 $\mathrm{ns}$ suggests that the moving Al atoms diffuse across the AC, as shown in the next subsection.

\begin{figure}[h]
 \begin{center}
\includegraphics[keepaspectratio,width=1 \linewidth]{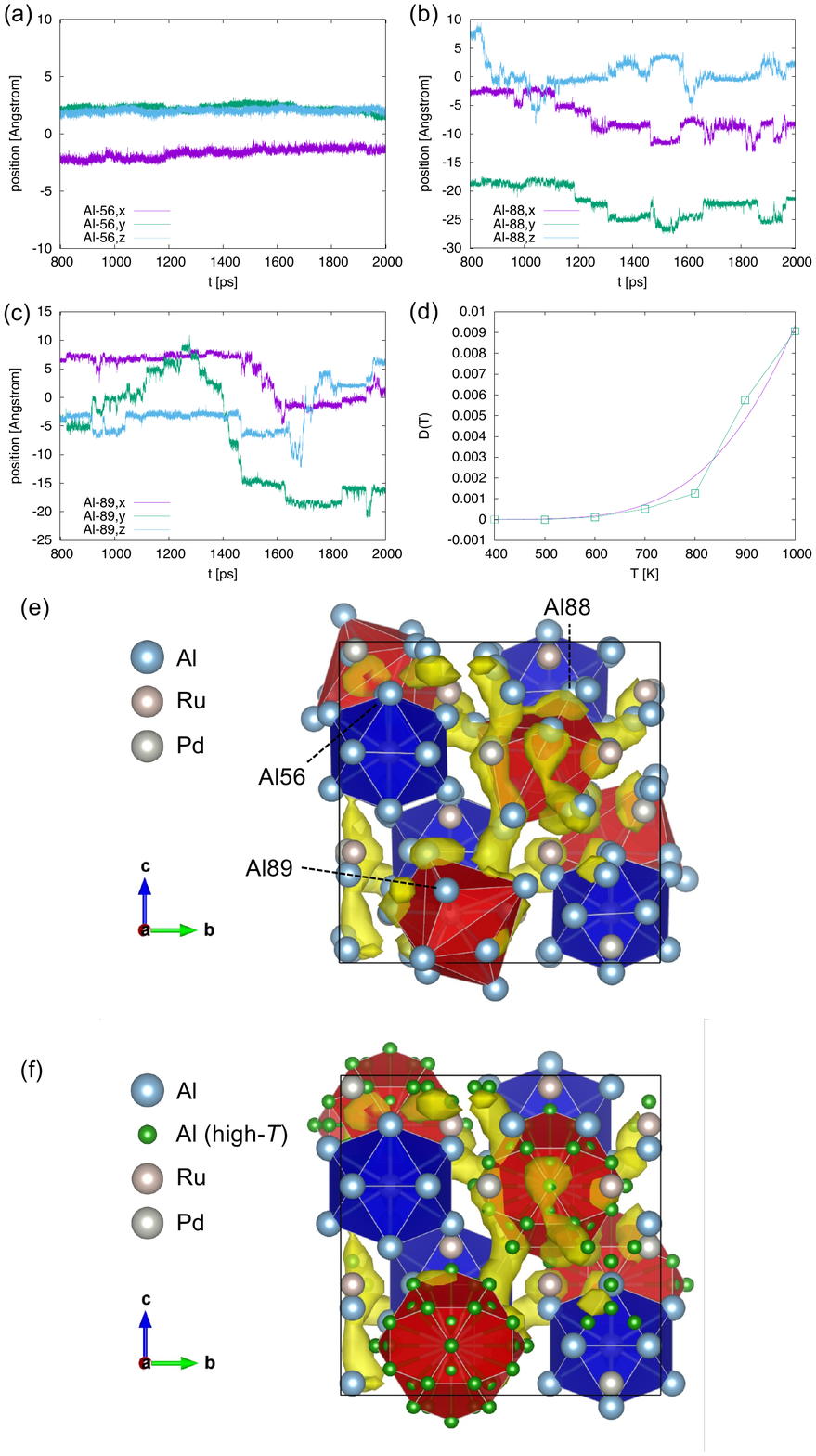}
\caption{\label{fig:phason} (a)--(c) Real-time dynamics of the Al atoms in the 1/1 AC Al$_{92}$Pd$_{20}$Ru$_{16}$ at $1000\,\mathrm{K}$. 
    (d) Temperature dependence of the diffusion coefficient.
    Visualization of the diffusive Al atoms in (e) the initial atomic structure and (f) atoms generated by the hyperatom in the six-dimensional space. Here, the yellow isosurface and the blue and red polyhedra represent the three-dimensional distribution of the Al atoms for 1/1 AC at $T = 1000\,\mathrm{K}$ from $t = 800\,\mathrm{ps}$ to $2000\,\mathrm{ps}$, and the inner shell of the mini-Bergman, and pseudo-Mackay clusters, respectively. }
\end{center}
\end{figure}

\subsection*{Atom diffusion}
Figure \ref{fig:phason}(d) shows the diffusion coefficients $D(T)$ of the Al atoms in 1/1 AC, which are obtained from the slope of the mean square displacement of the moving Al atoms; see Supplementary Material for the computational details. 
By a linear fitting of the Arrhenius plot of $D(T)$, the free energy barrier of the Al jumps is estimated as $\Delta E \sim 6000\,\mathrm{K}$ ($\sim 0.52\,\mathrm{eV}$), which is reachable at the high-temperature range exhibiting the anomalous heat capacity. 
Here it is worth noting that the diffusion of Al atoms in ACs (and presumably in QCs) occurs without any vacancy formation, as opposed to most diffusion mechanisms in perfect crystals. 
Figure \ref{fig:phason}(e) visualizes the trajectory of the Al diffusion. One can see that the diffusive paths are restricted within the locations at the edge of the mini-Bergman and pseudo-Mackay clusters shown in Figure \ref{fig:structure}.

\section*{Discussions} \label{sec:discussions}

\subsection*{Origin of Al diffusion}
\begin{figure}[tb]
 \begin{center}
\includegraphics[keepaspectratio,width=1 \linewidth]{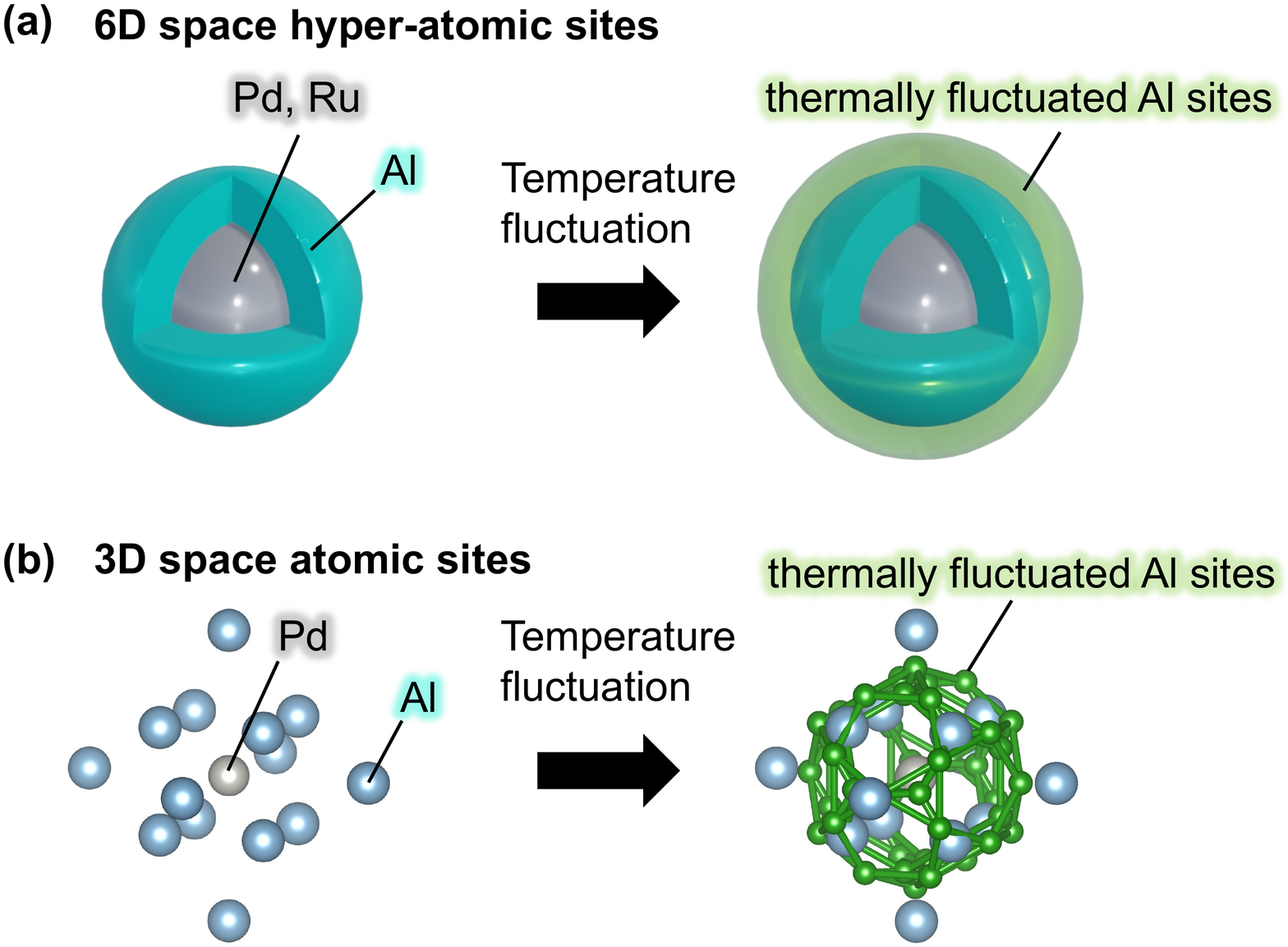}
\caption{\label{fig:superatom} (a) Schematic figure of the effect of temperature fluctuation of the hyperatom in six-dimensional space. Al atoms are located around the surface. (b) Schematic figure of the effect of a temperature fluctuation of the corresponding atoms composing the inner shell of the pseudo-Mackay cluster in three-dimensional space.}
\end{center}
\end{figure}

How can we understand the diffusion pathways? 
As mentioned earlier, the static configurations of the QCs and ACs correspond to the projection of hyperatoms on a periodic lattice in a higher dimensional space. 
Figures \ref{fig:superatom}(a) and 4(b) show the positions of the hyperatomic sites and atoms in the real space of the Al--Pd--Ru icosahedrons QC and AC, respectively. 
Here, the hyperatomic sites of the Pd and Ru atoms are located inside the hyperatomic shell, while the Al atoms are located outside. 
The hyperatomic sites shown in Figure \ref{fig:superatom}(a) are schematic. 
In this case, the QC and AC static configurations correspond to the most compactly occupied hyperatomic sites in six-dimensional space. 
However, there are many other accessible hyperatomic Al sites outside the shell. 
The accessibility of each of these sites is energy-dependent and can only be determined after MD simulations. 
The movement between the different hyperatomic sites corresponds to a phason flip, and a series of phason flips were observed at finite temperatures in the MD simulations. 
At higher temperatures, the occupancy of Al hyperatomic sites beyond the outer shell increases, resulting in the Al coordination shown in Figure \ref{fig:superatom}(b). 
This corresponds exactly to the diffusion path of Al atoms as found in the MD simulation shown in Figure \ref{fig:phason}(f). 
Therefore, the Al diffusion in Al--Pd--Ru QC and AC can be regarded as a six-dimensional hyperatomic fluctuation.

\subsection*{Origin of anomalous heat capacity}
Why does the heat capacity become anomalous in the presence of phason flips? 
The heat capacity follows the Dulong--Petit rule assuming harmonic vibrations of conventional phonons. 
The anharmonicity of the potential energy surface is the origin of the anomaly. 
According to QC hydrodynamics, phason flips correspond to diffusive Nambu--Goldstone modes\cite{10.21468/SciPostPhys.9.5.062,PhysRevResearch.2.022022}. 
Their dispersion relation is expressed as 
\begin{align}
\omega = -i D(T) q^2 + \cdots,
\end{align}
where $q$ denotes the momentum of the phason mode.
In this case, the imaginary frequency of the diffusive mode, $i\omega$, is proportional to the diffusion coefficient $D(T)$.
The magnitude of the heat capacity is determined by how much energy a material absorbs when heat is applied from the outside. 
Because the applied heat is partially used to excite the diffusive Nambu--Goldstone mode, the heat capacity becomes larger with increasing temperature. 

\section*{Summary}
The anomalous heat capacity observed in icosahedral AC in the Al--Pd--Ru system was well reproduced by machine-learning MD simulations. 
Some of the Al atoms were found to diffuse through the crystal in the absence of vacancies via nearly discontinuous jumps that can be regarded as phason flips. 
The restricted pathways of the phason flips can be understood as thermal fluctuations of hyperatoms in six-dimensional space. 
From this result, we conclude that anomalous heat capacity is caused by atomic diffusion due to thermal fluctuations of hyperatoms in the higher-dimensional space in QCs and ACs.
This study suggests that the high dimensionality of QC structure may affect physical properties other than heat capacity.

\subsection*{Acknowledgements}

Y.N. was partially supported by JSPS KAKENHI Grant Numbers 18K11345, 20H05278, 22H04602, and 22K03539. The calculations were performed using the supercomputing system HPE SGI8600 at the Japan Atomic Energy Agency.
Y.I. was supported by JSPS KAKENHI under Grant 19H05818, 19H05821, and 19J21779.
M.S. was supported by JSPS KAKENHI (18H05519, 21H01603) and the Fugaku Battery \& Fuel Cell Project. 
We thank Edanz (https://jp.edanz.com/ac) for editing a draft of this manuscript.

Y.N. and Y.I. contributed equally to this work.

\bibliography{bib}

\begin{thebibliography}{41}
\bibitem [{not()}]{note}%
  \BibitemOpen
  \href@noop {} {}\bibinfo {note} {The ANN fingerprint parameters were set as
  radial\_Rc = 8.0, radial\_N = 16, angular\_Rc = 6.5, and angular\_N = 4 for
  all the atomic elements. We used two hidden layers with 15 units and the
  tanh-type activation function.}\BibitemShut {Stop}%
\end{thebibliography}

\pagebreak

\section*{Supplemental materials}

\subsection*{Methods}
\subsubsection*{Sample preparation}
 Synthesis of Al--Pd--Ru QC, 2/1 AC, 1/0 AC was performed from commercial element powders, Al (3$N$ purity; Kojundo Chemical Laboratory Co., Ltd., Japan), Pd (3$N$ purity; TANAKA Kikinzoku Kogyo K.K, Japan), and Ru (3$N$ purity; TANAKA Kikinzoku Kogyo K.K, Japan) with nominal compositions of Al$_{71.5}$Pd$_{19}$Ru$_{9.5}$ for QC, Al$_{71}$Pd$_{20}$Ru$_{9}$ for 2/1 AC, Al$_{68}$Pd$_{20.5}$Ru$_{11.5}$ for 1/0 AC, respectively.
The mixed powders were pressed under $200\,\mathrm{MPa}$ into disc-shaped pellets. Each pellet was melted in an arc-melting furnace under an argon atmosphere.
The ingots wrapped in tantalum foil were annealed at $1173\,\mathrm{K}$ for $72\,\mathrm{h}$ under argon atmosphere.
The annealed ingots were reground into powder and then sintered by spark plasma sintering.
The temperatures of the specimens were increased from room temperature to $1173\,\mathrm{K}$, and then the samples were kept for $20\,\mathrm{min}$ at a uniaxial pressure of $90\,\mathrm{MPa}$ under an argon atmosphere.
The sintered samples were annealed again under the same conditions as above.

\subsubsection*{Characterization}
 Phase identification and determination of lattice constants were performed by the powder X-ray diffraction (XRD) method with Cu~K--L$_{2,3}$ radiation (MiniFlex; Rigaku Co., Japan). Each XRD pattern is shown in the supplemental materials.
The longitudinal and transverse speeds of sound $v_{\mathrm{l}}$ and $v_{\mathrm{t}}$ were measured by the ultrasonic pulse-echo method using an echometer (Echometer 1062; Karl Deutsch Co., Germany).
The heat capacity at constant pressure $C_p$ was measured by differential scanning calorimetry using a differential scanning calorimeter (DSC 404 F1 Pegasus; NETZSCH Japan K.K.).

\subsubsection*{Physical properties analysis}
$C_{V}/k_{\mathrm{B}}$ is converted by using the following thermodynamic relation:
\begin{equation}
  C_{V}=(1+9VB_S\alpha^2TC_{p}^{-1})^{-1}C_{p}
\end{equation}
where $V$, $B_S$, and $\alpha$ are the average volume per atom, the adiabatic bulk modulus, and the linear thermal expansion coefficient, respectively.
$B_S(300\,\mathrm{K})$ is calculated by using the density $\rho$, $v_{\mathrm{l}}$ and $v_{\mathrm{t}}$ as $B_S=\rho(v_{\mathrm{l}}^2-\frac{4}{3}v_{\mathrm{t}}^2)$.
$B_S(300\,\mathrm{K})/B_S(T)$ where $T>300\,\mathrm{K}$ for Al--Pd--Ru QC and the ACs are assumed to be the same as that of Al--Pd--Mn QC measured by Tanaka \textit{et al.}\cite{Tanaka1996}.
Here, $B_S(T)$ is fitted to the function $B_S = B_0 - AT \exp{(-\frac{T_0}{T})}$\cite{Wachtman1961} with $B_0=127\,\mathrm{GPa}$, $A=0.044\,\mathrm{GPa}/\mathrm{K}$ and $T_0=486\,\mathrm{K}$.
$\alpha$ is extracted from the previously reported value for Al--Pd--Mn QC measured by Fukushima \textit{et al.} \cite{Fukushima_2021}.
We note that the trend of $\alpha$ is almost independent of chemical composition and the degree of approximation according to previous work \cite{EDAGAWA2001293}.
The maximum difference between $C_{p}/k_{\mathrm{B}}$ and $C_{V}/k_{\mathrm{B}}$ is up to 0.3$k_{\mathrm{B}}$ at $T=1050\,\mathrm{K}$ in QCs.


\subsubsection*{Simulation setups}
 We consider a $2 \times 2 \times 2$ supercell of the 1/0 AC Al$_{23}$Pd$_{2}$Ru$_{6}$, the 1/1 AC Al$_{92}$Pd$_{20}$Ru$_{16}$, and the 2/1 AC Al$_{396}$Pd$_{80}$Ru$_{68}$. 
We also consider Al$_{13}$Ru$_{4}$ as a reference material where the heat capacity is similar to the Dulong--Petit limit and has weak temperature dependence. 
We construct artificial neural networks (ANNs) to reproduce the energy calculated by the DFT calculations. 
The DFT-MD, ML-MD and SLHMC were implemented in \texttt{PIMD} software \cite{PIMD, PhysRevB.102.041124,doi:10.1063/5.0055341}, 
which supports the interface of both the Vienna Ab initio Simulation Package (\texttt{VASP}) \cite{PhysRevB.47.558,PhysRevB.54.11169}
and Atomic Energy Network (\texttt{aenet}) software \cite{ARTRITH2016135}. 
The DFT calculations were conducted using the Perdew--Burke--Ernzerhof functional \cite{PhysRevLett.77.3865}.
The projector augmented wave method \cite{PhysRevB.50.17953,PhysRevB.59.1758} was employed, while the cutoff energy was $500\,\mathrm{eV}$ and the sampling point was only the $\Gamma$ point. 
In \texttt{aenet}, we adopt the Chebyshev basis set as a descriptor for atomic environments \cite{PhysRevB.96.014112}, and the corresponding parameters are given in Ref.~\onlinecite{note}. 
SLHMC is used for improving the ANN potentials \cite{PhysRevB.102.041124,doi:10.1063/5.0055341}.
The heat capacity is calculated by the NVT ensemble average of the energy fluctuation. 
The massive Nosé--Hoover chain method is used for generating the NVT ensemble for heat capacity calculations.
In calculations of the mean-square displacement and activation energy, the Nosé--Hoover thermostat is used to study real-time dynamics. 
Using the $1\,\mathrm{fs}$ time step, we do simulations with up to $2 \times 10^6\,\mathrm{fs}$.
The quality of the ANN potentials is confirmed with SLHMC calculations (see supplemental materials) because the SLHMC generates the same ensemble as that generated by the first-principle MD simulation \cite{PhysRevB.102.041124}.

\subsection*{Quality of the machine-learning potential}
\begin{figure*}[th]
  \begin{center}
(a)\begin{minipage}[b]{0.4\linewidth}
        \centering
        \includegraphics[keepaspectratio,width=\linewidth]{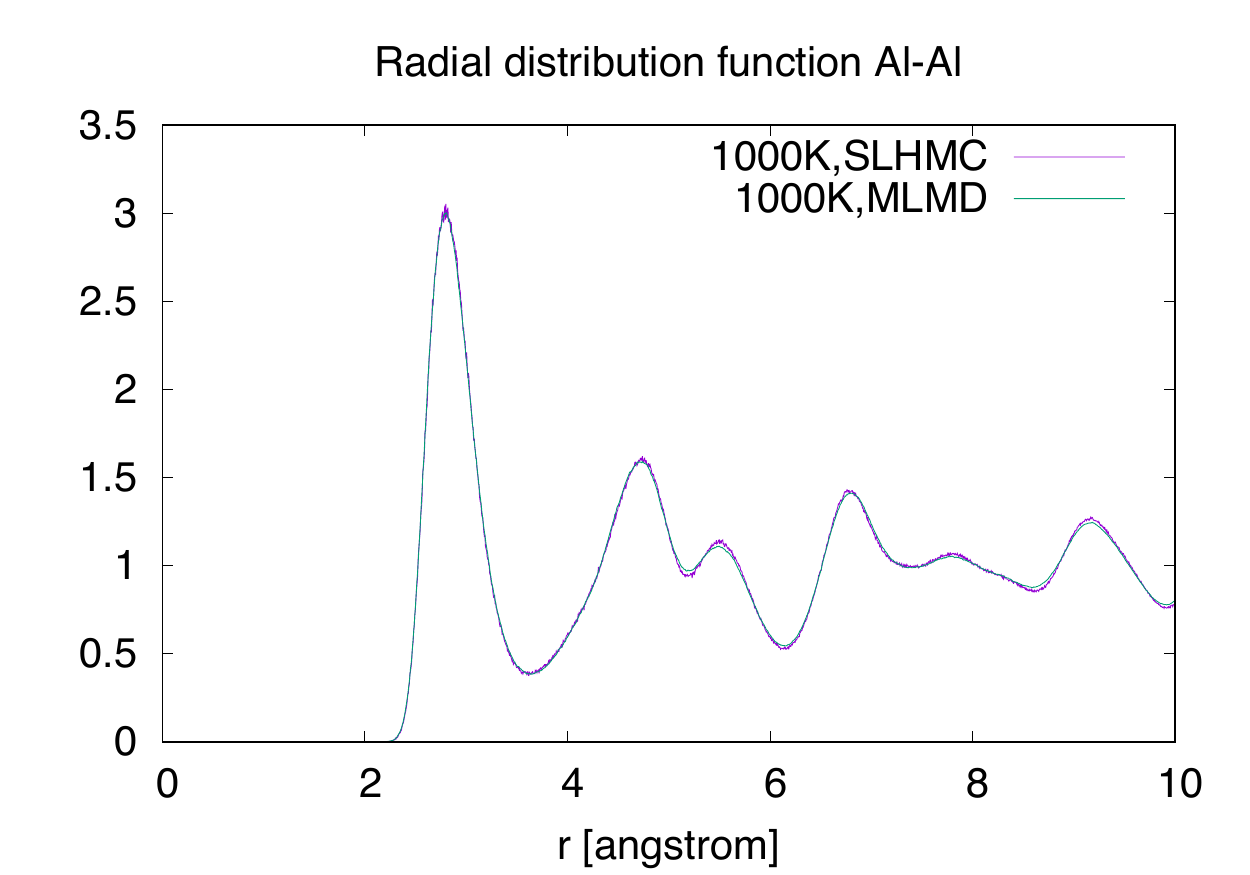}
    \end{minipage}
(b)\begin{minipage}[b]{0.4\linewidth}
        \centering
        \includegraphics[keepaspectratio,width=\linewidth]{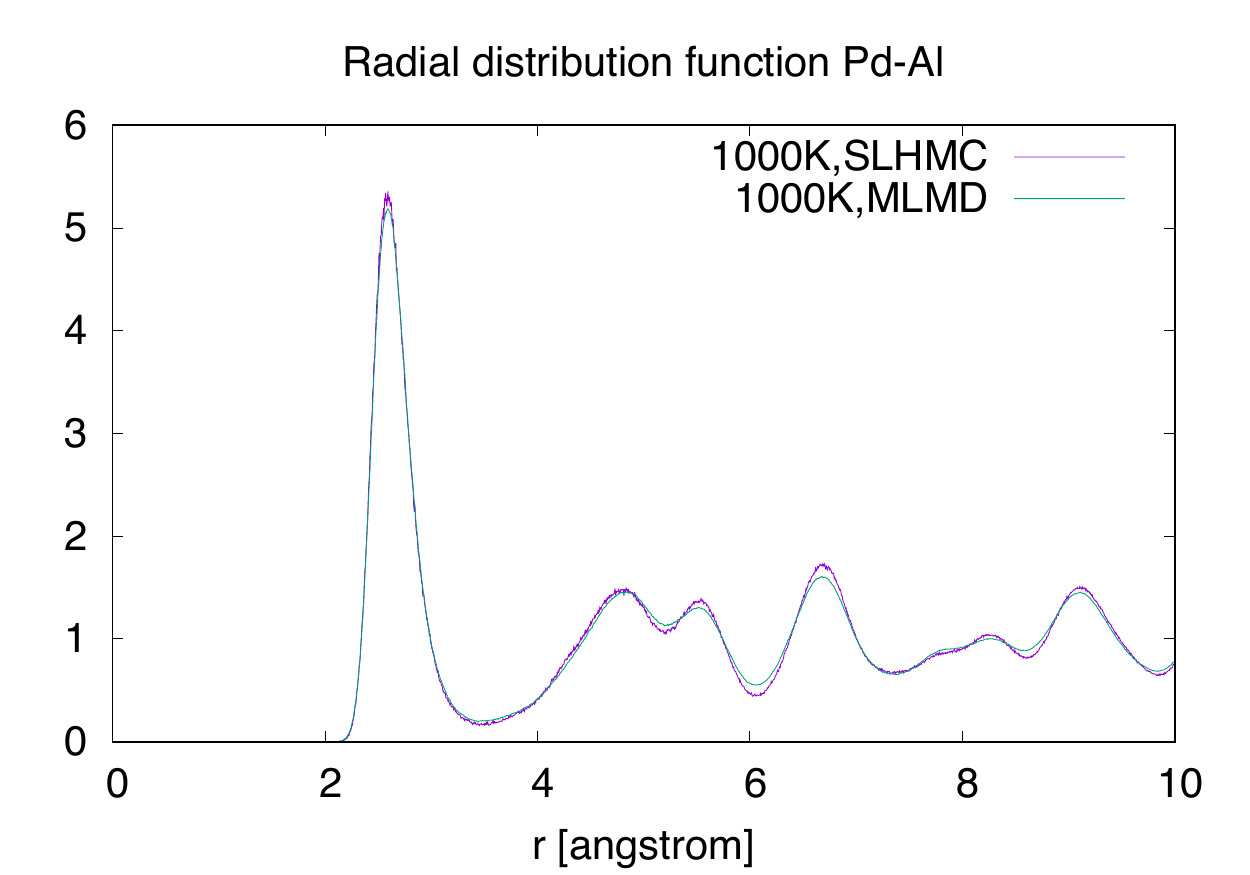}
    \end{minipage}\\
(c)\begin{minipage}[b]{0.4\linewidth}
        \centering
        \includegraphics[keepaspectratio,width=\linewidth]{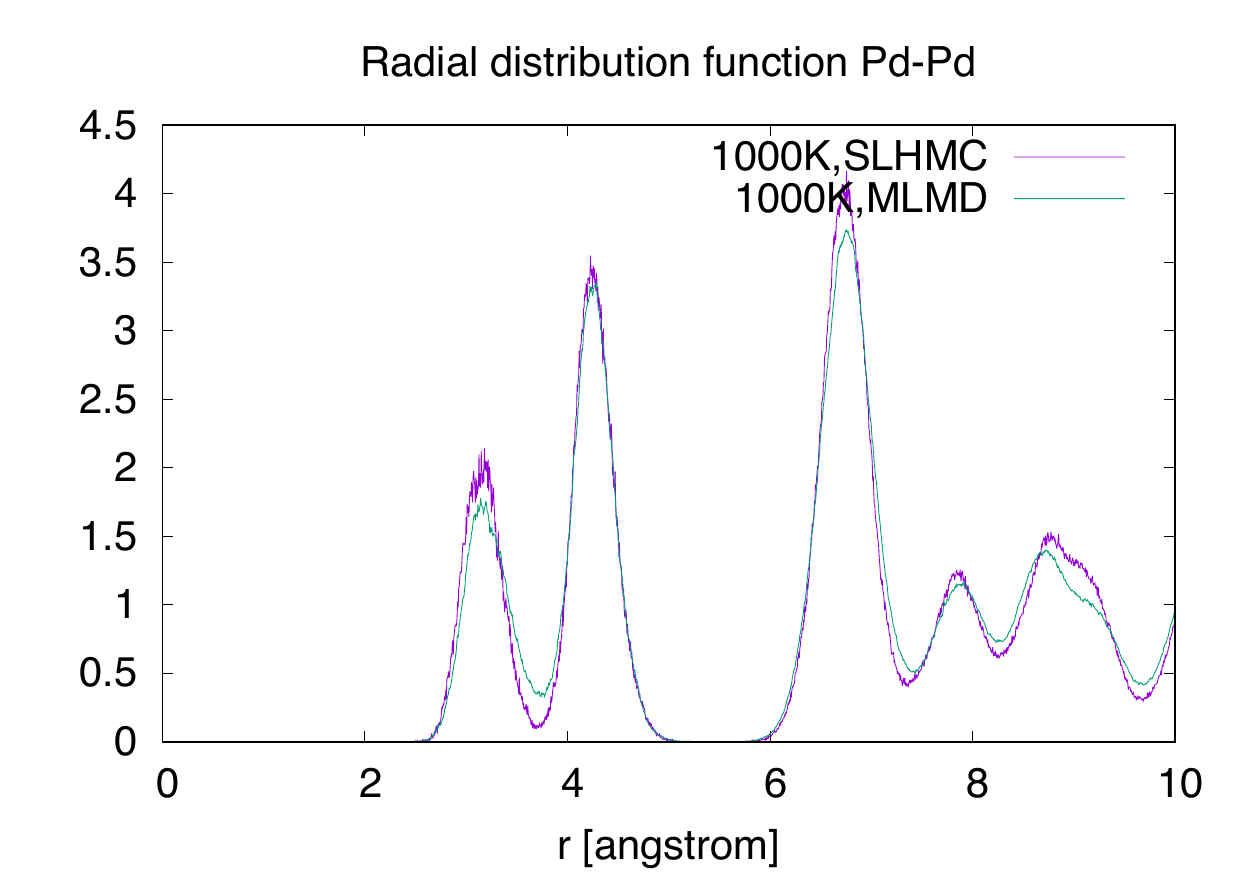}
    \end{minipage}
(d)\begin{minipage}[b]{0.4\linewidth}
        \centering
        \includegraphics[keepaspectratio,width=\linewidth]{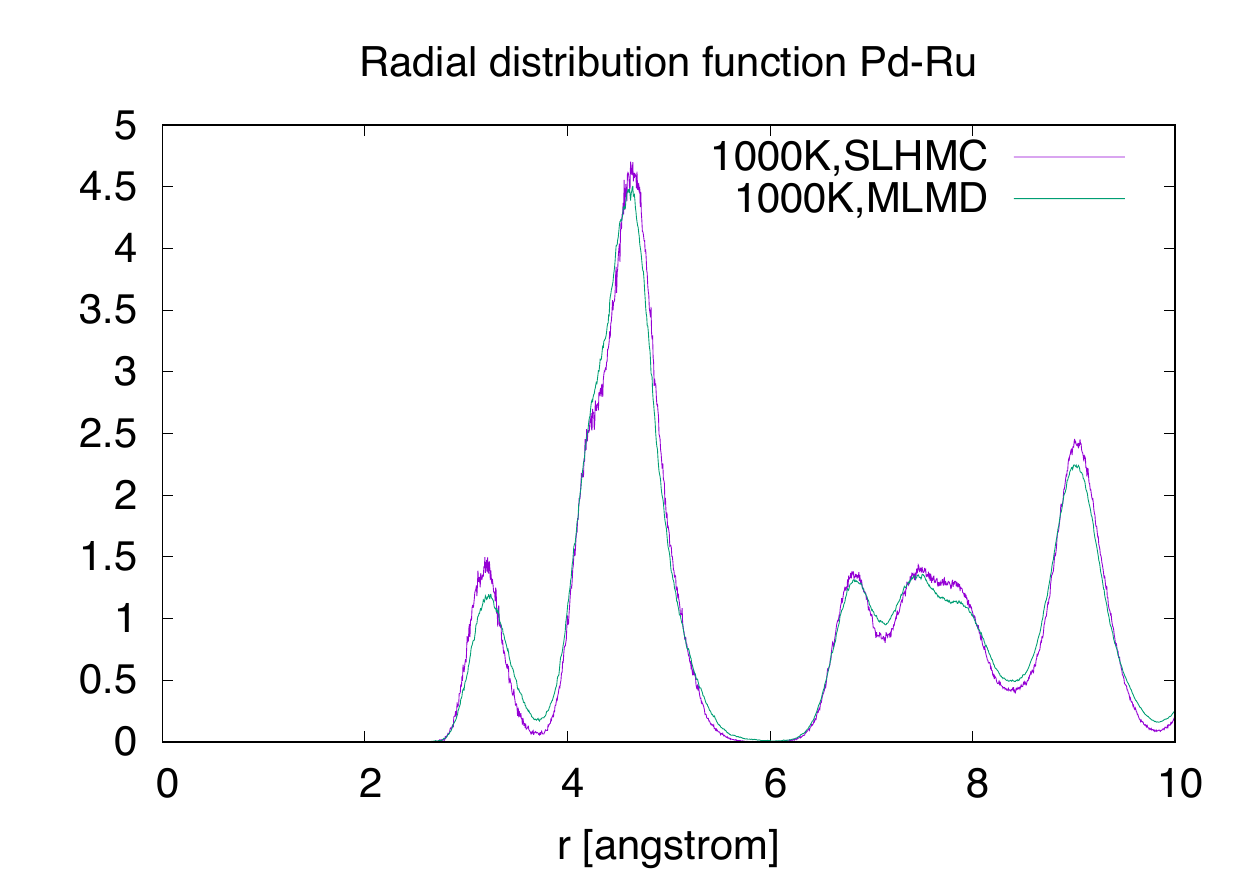}
\end{minipage}     \\
(e)\begin{minipage}[b]{0.4\linewidth}
        \centering
        \includegraphics[keepaspectratio,width=\linewidth]{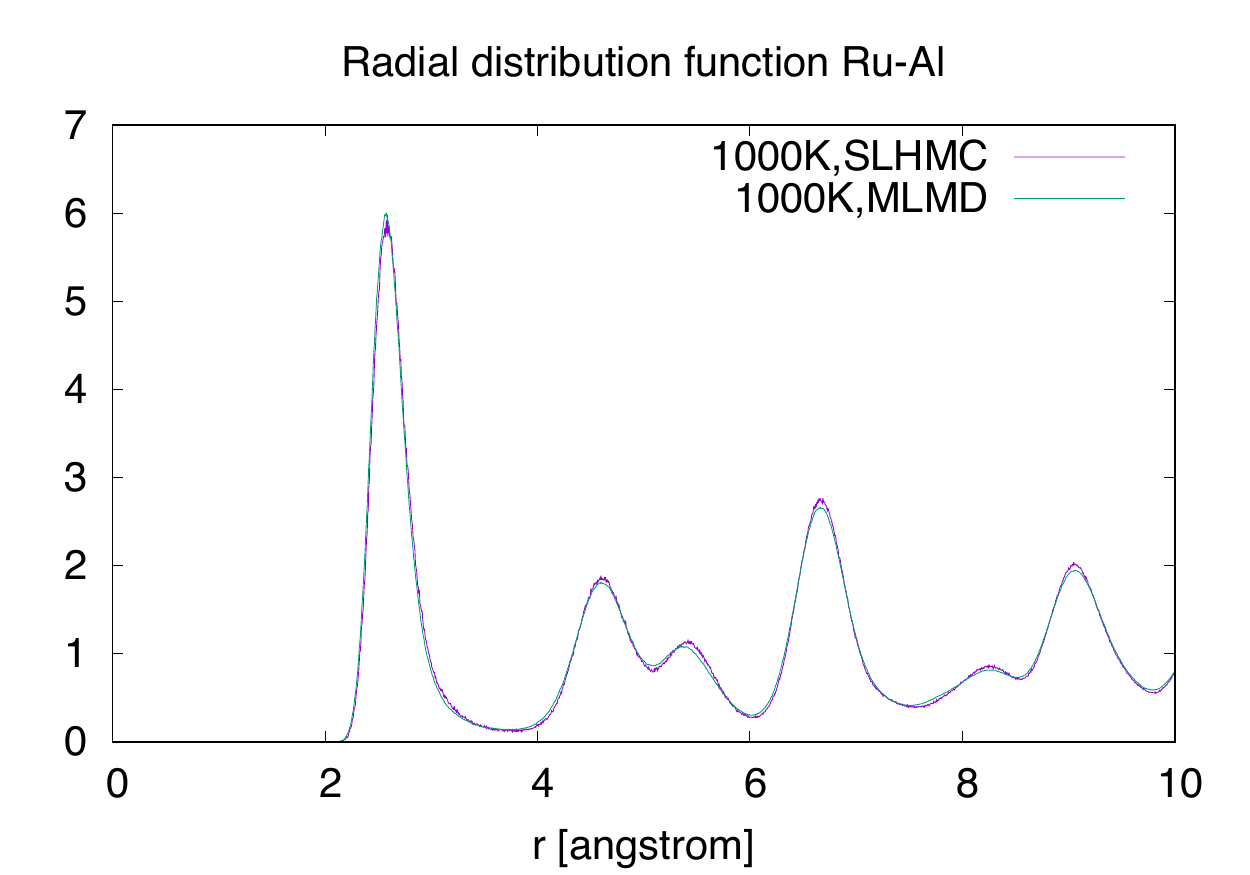}
    \end{minipage}
(f)\begin{minipage}[b]{0.4\linewidth}
        \centering
        \includegraphics[keepaspectratio,width=\linewidth]{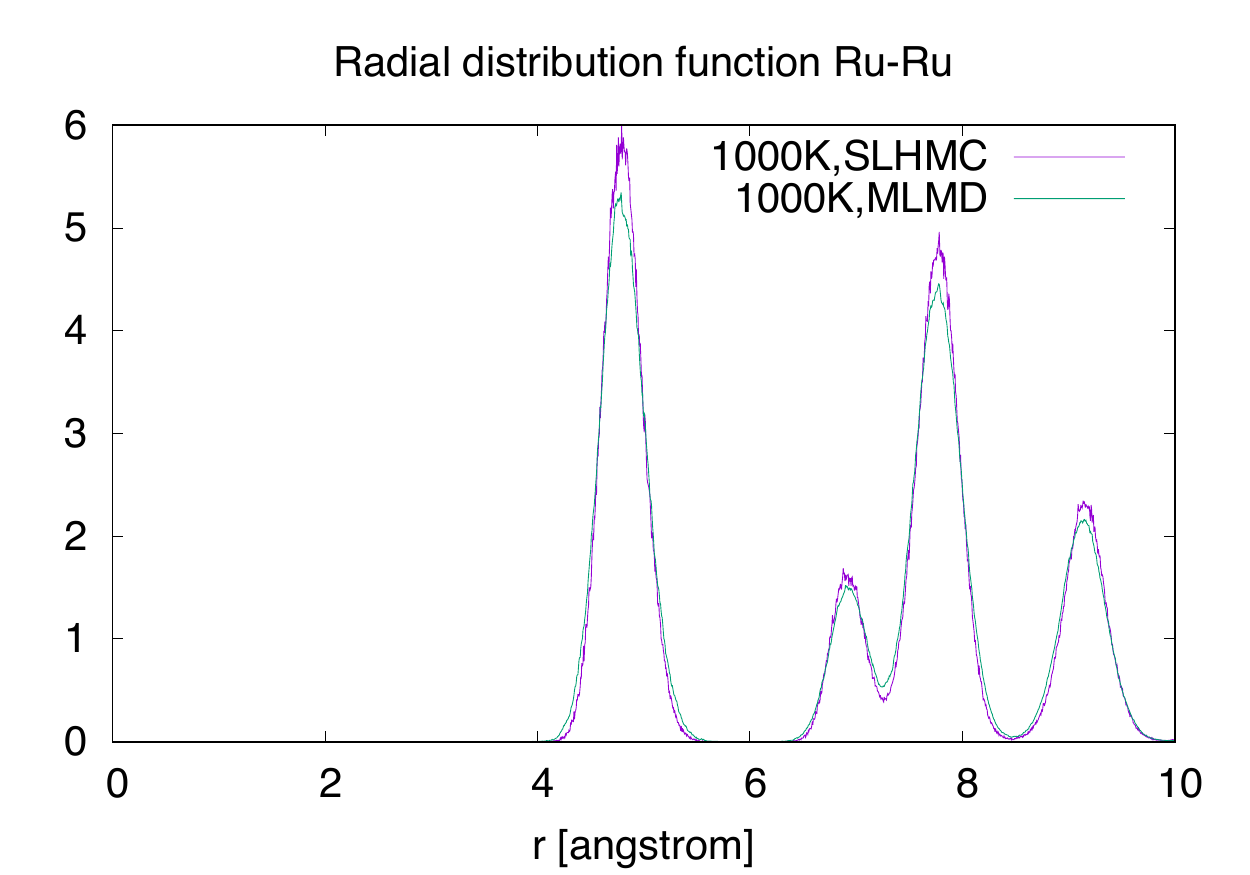}
\end{minipage}     \\
    \caption{\label{fig:rdf} Radial distribution functions for 1/1 approximant.
    }
    \label{fig:phason}
    \end{center}
\end{figure*}

The quality of the machine-learning potential can be checked with the self-learning hybrid Monte Carlo method (SLHMC). 
The ensemble generated by the SLHMC is equivalent to that of the DFT calculations. 
As shown in Fig.~\ref{fig:rdf}, the radial distribution functions obtained in the MLMD simulations are similar to that obtained in the SLHMC simulations.
This ensures that our machine learning potential performs close to the accuracy of DFT and is sufficient to study the physical properties of ACs.

\subsection*{Mean square displacement}
\begin{figure}[th]
 \begin{center}
\includegraphics[keepaspectratio,width= 0.7 \linewidth]{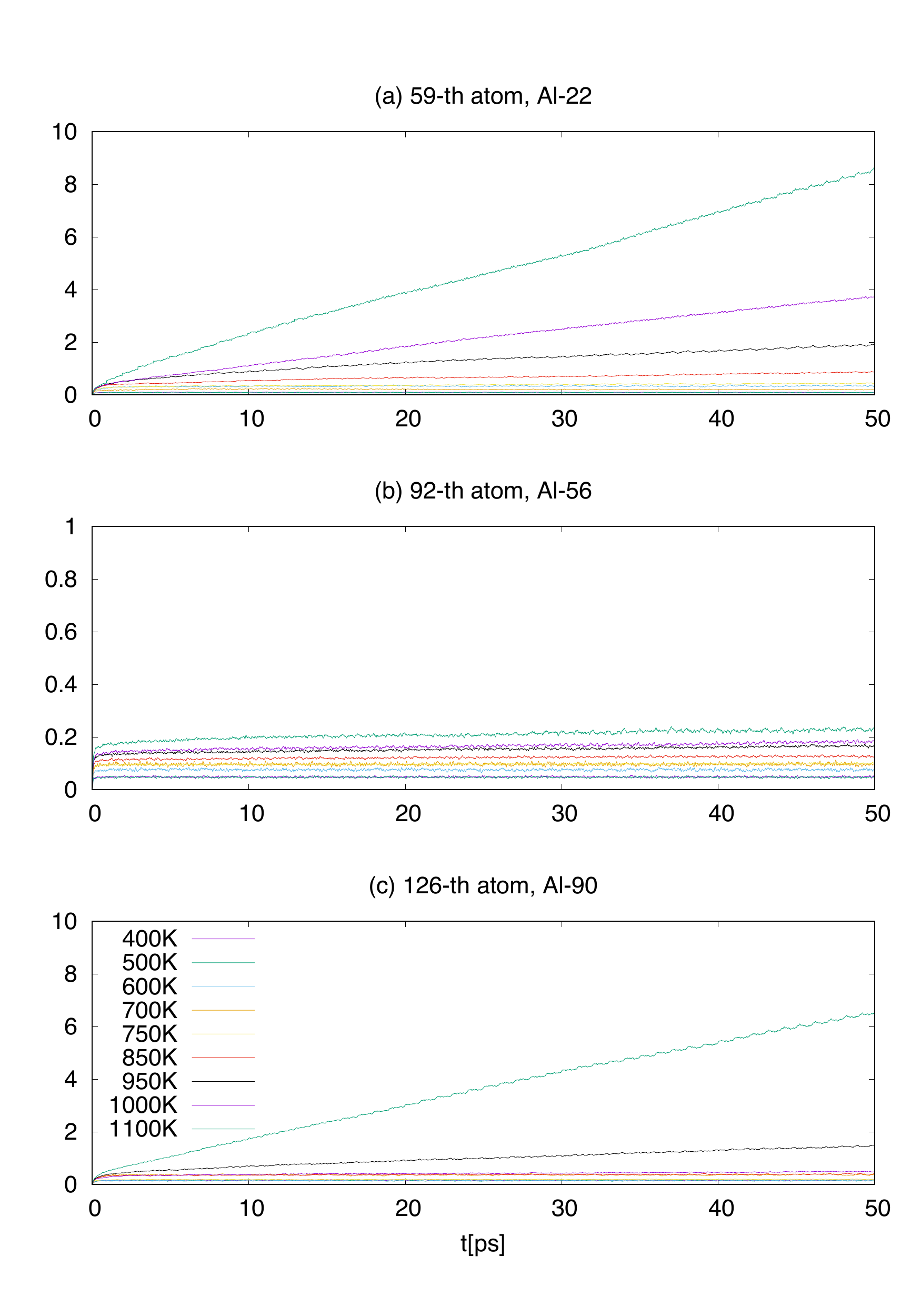}
\caption{\label{fig:mse} (Color online) Mean square displacement of (a) Al-22, (b) Al-56, and (c) Al-90 atoms for 1/1 approximant. }
\end{center}
\end{figure}

The mean square displacement is defined as 
\begin{align}
    {\bf MSD}(t) = \frac{1}{N} \sum_{j=1}^N \phi_j({\bm r}_j(t)), 
\end{align}
where 
\begin{align}
     \phi_j({\rm r}_j(t)) \equiv \langle [{\bm r}_j(t) - {\bm r}_j(0)]^2 \rangle.
\end{align}
We call $\phi_j({\bm r}_j(t))$ the MSD for $j$-th atom. 
As shown in Fig.~\ref{fig:mse}, some of Al atoms become diffusive with increasing the temperature. 
We calculate the temperature dependence of the diffusion coefficient $D(T) \equiv (1/6) \lim_{t \rightarrow \infty}{\rm MSD}(t)/t $. 
We can fit the diffusion coefficient by an Arrhenius law, 
\begin{align}
    D(T) = D_0 \exp \left( -\Delta E/ k_{\rm B} T \right).
\end{align}

\section*{X-ray diffraction patterns}

\begin{figure}[th]
  \begin{center}
\includegraphics[keepaspectratio,width=1 \linewidth, page=1]{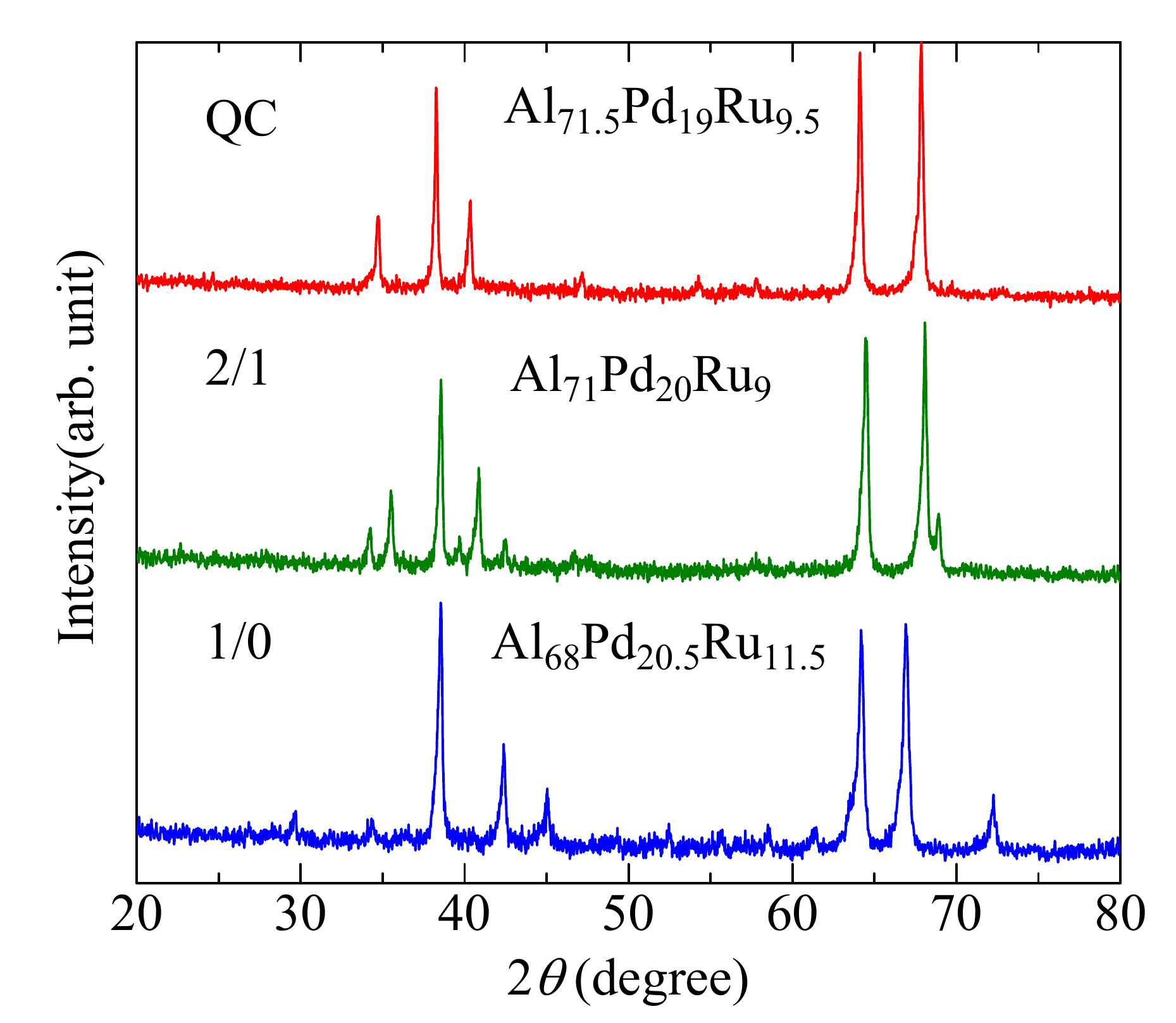}
    \caption{\label{fig:xrd} (Color online) The X-ray diffraction patterns of Al-Ru-Pd QC (RED), 2/1 AC (GREEN), and 1/0 AC (BLUE), respectively. }
    \end{center}
\end{figure}

Figure \ref{fig:xrd} shows the experimental X-ray diffraction patterns of Al-Ru-Pd quasicrystal, 2/1 approximant, and 1/0 approximant.

\end{document}